\documentclass{ws-procs9x6}

\begin{document}

\title{Nucleon Form Factors in Point-Form Spectator-Model Constructions}

\author{T. Melde}

\address{Theoretische Physik, Institut f\"ur Physik,
Karl-Franzens-Universit\"at, \\
Universit\"atsplatz 5, A-8010 Graz, Austria}

\begin{abstract}
We discuss electromagnetic currents in the point-form formulation of 
relativistic quantum mechanics. The construction is along a spectator
model and implies that only one quark is explicitly coupled to the
photon, but  nevertheless many-body contributions are present
in the current operator. Such effects are unavoidable in relativistic 
constructions and resulting ambiguities are notably reduced by 
imposing charge normalization and time-reversal invariance.
The residual theoretical indetermination introduces small but 
sizeable changes in the nucleon 
form-factors, particularly at higher $Q^2$ values, with the data generally 
centered in the middle of the theoretical band. 
\end{abstract}


\bodymatter

\section{Introduction}\label{aba:sec1}
Nucleon form factors are of considerable importance,
as they can give additional information on the internal structure of the nucleons. 
Renewed interest in the form factors is due to new polarization transfer 
measurements~\cite{Jones:1999rz,Gayou:2001qd,Punjabi:2005wq,%
MacLachlan:2006vw} for the proton, that produced data at variance with respect
to earlier Rosenbluth 
data~\cite{Andivahis:1994rq,Walker:1989af,Hohler:1976ax}. 
Accurate
measurements~\cite{Christy:2004rc,Qattan:2004ht} of new generation revealed
the existence of a discrepancy that cannot be 
explained~\cite{Arrington:2004is} with Coulomb distortion effects.
It has been suggested that  two-photon contributions could be
the source of the disagreement~\cite{Arrington:2003qk,Arrington:2004ae},
but to date the experimental situation is still under discussion (see 
also~\cite{Tvaskis:2005ex,Jones:2006kf,Tomasi-Gustafsson:2006pa%
}).
On the theoretical side, two-photon 
corrections~\cite{Guichon:2003qm,Blunden:2003sp,Blunden:2005ew,%
Chen:2004tw,Afanasev:2005mp}
and additional $\Delta$ contributions~\cite{Kondratyuk:2005kk}
have been advocated to resolve this problem. A detailed discussion on
the theoretical and experimental status of nucleon form factors can be found 
in two recent reviews~\cite{Arrington:2006zm,Perdrisat:2006hj}.

Here, we reinvestigate the description of nucleon form factors within the
Goldstone-boson exchange (GBE) constituent-quark model 
(CQM)~\cite{Glozman:1998ag}. The approach is complementary to field-theoretic
ones, like e.g. Refs.~\cite{Oettel:2000jj,Alkofer:2004yf}, 
as it is valid for low and intermediate momentum transfers. The electromagnetic 
(EM) ground state properties of the nucleon already were reproduced quite 
well~\cite{Wagenbrunn:2000es,Glozman:2001zc,Boffi:2001zb,Berger:2004yi}
with the point-form spectator model (PFSM) applied to the GBE CQM. 
With the advent of newer and more precise data for the nucleon form-factor,
it becomes necessary to assess more carefully the predictive power of
the PFSM construction and possible sources of 
indetermination~\cite{Melde:2006jn}. 
In the last Cortona meeting~\cite{Melde:2004ra}, we presented
arguments that translational invariance of the transition amplitude implies the
presence of many-body contributions. In Ref.~\cite{Melde:2004qu} it has been shown
that in connection to these many-body effects certain ambiguities arise which
cannot be fully constrained by 
Poincar\'e invariance alone. Here, we further constrain the PFSM construction
by requiring charge normalization and time-reversal invariance of the elastic
electromagnetic nucleon transition amplitude.
\def\figsubcap#1{\par\noindent\centering\footnotesize(#1)}

\begin{figure}[t]
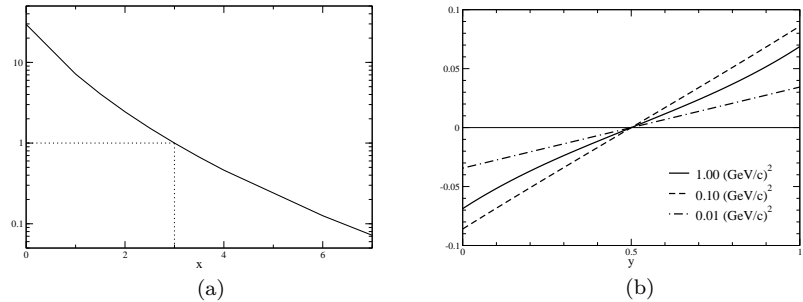
%
\begin{center}
  \parbox{2.1in}{\epsfig{figure=figure_1.eps,width=1.9in}\figsubcap{a}}
  \hspace*{4pt}
  \parbox{2.1in}{\epsfig{figure=figure_2.eps,width=1.9in}\figsubcap{b}}
  \caption{ 
  (a) Proton electric form factor $G_E^p$ at momentum transfer $Q^2=0$ (i.e. proton charge) 
as a function of the exponent $x$ in the normalization factor
$\cal N$ of Eq.~(\ref{eq:offsymfac}). 
(b) Expectation value of the electromagnetic current component $\hat J^{\mu=3}$
in the Breit frame as a function of the exponent $y$ in the normalization factor
$\cal N$ of (\ref{eq:offsymfac}) for three different values of the momentum
transfer $Q^2$.
}%
  \label{fig1}
\end{center}
\end{figure}
\def\figsubcap#1{\par\noindent\centering\footnotesize(#1)}
\begin{figure}[t]
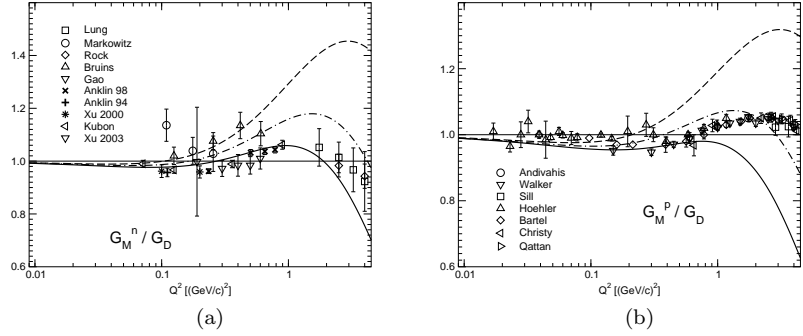
%
\begin{center}
  \parbox{2.1in}{\epsfig{figure=figure_3a.eps,width=1.9in}\figsubcap{a}}
  \hspace*{4pt}
  \parbox{2.1in}{\epsfig{figure=figure_3b.eps,width=1.9in}\figsubcap{b}}
  \caption{ Comparison of magnetic form-factor to dipole ratios for the neutron (a) 
  and proton (b) in the GBE CQM with PFSM currents.
The full line denotes the theoretical results with ${\cal N}_{\rm geo}$, the dashed line for
${\cal N}_{\rm ari}$ and the dash-dotted line for ${\cal N}_{\rm fit}$. All ratios are normalized to
one at $Q^2=0$.
Experiment given 
by Refs.~\protect\cite{Lung:1993bu,Markowitz:1993hx,%
Rock:1982gf,Bruins:1995ns,Gao:1994ud,Anklin:1994ae,Anklin:1998ae%
,Xu:2000xw,Kubon:2001rj,Xu:2002xc} 
and Refs.~\protect\cite{Andivahis:1994rq,Walker:1989af,Sill:1993qw,Hohler:1976ax,Bartel:1973rf,%
Christy:2004rc,Qattan:2004ht}
}%
  \label{fig2}
\end{center}
\end{figure}
\def\figsubcap#1{\par\noindent\centering\footnotesize(#1)}
\begin{figure}[t]
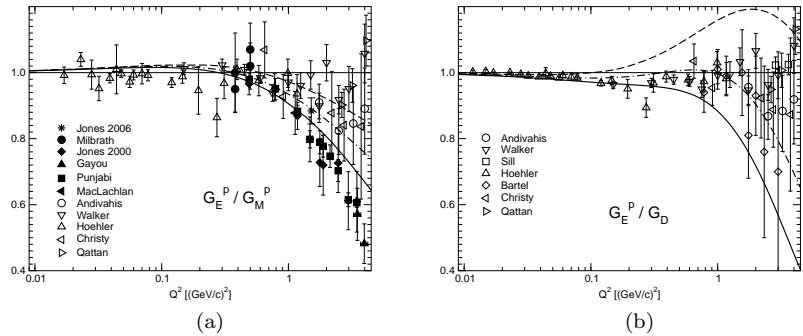
%
\begin{center}
  \parbox{2.1in}{\epsfig{figure=figure_6a.eps,width=1.9in}\figsubcap{a}}
  \hspace*{4pt}
  \parbox{2.1in}{\epsfig{figure=figure_6b.eps,width=1.9in}\figsubcap{b}}
  \caption{Notation same as~\fref{fig2} but for the proton form factor ratios; 
  electric to magnetic (a) as well as electric to dipole (b). 
Experimental data from 
Refs.~\protect\cite{%
Andivahis:1994rq,Walker:1989af,Sill:1993qw,Hohler:1976ax,Bartel:1973rf,%
Christy:2004rc,Qattan:2004ht, %
Jones:2006kf,Milbrath:1998de,Jones:1999rz,Gayou:2001qd,Punjabi:2005wq,%
MacLachlan:2006vw%
}.
}
  \label{fig3}
\end{center}
\end{figure}
\section{The Electromagnetic Current}
The reduced matrix element of the PFSM EM current
between three constituent quarks with individual
momenta $p_i$ and helicities $\sigma_i$ reads~\cite{Melde:2004qu}
\begin{eqnarray}
\left<p'_1,p'_2,p'_3;\sigma'_1,\sigma'_2,\sigma'_3\right|
{\hat J}^{\mu}_{\rm rd}
\left|p_1,p_2,p_3;\sigma_1,\sigma_2,\sigma_3\right>
=
\nonumber\\
3 {\cal N}
e_1{\bar u}\left(p'_1,\sigma'_1\right)
\gamma^\mu
u\left(p_1,\sigma_1\right) 
2p_{20}\delta^3\left({\vec p}_2-{\vec p}'_2\right)
2p_{30}\delta^3\left({\vec p}_3-{\vec p}'_3\right)
 \delta_{\sigma_{2}\sigma'_{2}}
   \delta_{\sigma_{3}\sigma'_{3}}
\label{eq:pfsm}\, .
\end{eqnarray}
For the nucleon electromagnetic form-factors the  transition
amplitude between the incoming and outgoing baryon eigenstates is given
by the following definition
\begin{equation}
F^{\mu}_{\Sigma',\Sigma}\left(Q^2\right)
=
\left<V',m_{\rm N},\frac{1}{2},\Sigma'\right|{\hat J}^{\mu}_{\rm rd}
\left|V,m_{\rm N},\frac{1}{2},\Sigma\right>\, .
\end{equation}
In the Breit frame, the nucleon Sachs form-factors are related to the transition amplitude 
$F^{\mu}_{\Sigma',\Sigma}$ according to the expressions
\begin{eqnarray}
F^{0}_{\Sigma',\Sigma}\left(Q^2\right)
&=&2m_{\rm N} G_{\rm E}\left(Q^2\right)\delta_{\Sigma'\Sigma}
\\
{\vec F}_{\Sigma',\Sigma}\left(Q^2\right)
&=&
iQ G_{\rm M}\left(Q^2\right)
\chi^\dagger_{\Sigma'}\left({\vec \sigma} \times {\vec e}_{z}\right)\chi_\Sigma^{\phantom\dagger}
\, ,
\end{eqnarray}
where $\Sigma', \Sigma=\pm \frac{1}{2}$ are the spin projection along the $z$-axis
of the nucleon and $\chi$ the corresponding Pauli spinors.

As has been observed in Ref.~\cite{Melde:2004qu},  
a normalization factor ${\cal N}$ appears in
the PFSM current of (\ref{eq:pfsm}), which can be parametrized in 
the following way
\begin{equation}
{\cal N}\left(x,y\right)=
\left(\frac{M}{\sum_i{\omega_{i}}}\right)^{xy}
\left(\frac{M'}{\sum_i{\omega'_{i}}}\right)^{x\left(1-y\right)}\, .
\label{eq:offsymfac}
\end{equation}
In this form $x$ and $y$ are 
free parameters which we consider in the range $0\le x,0\le y\le 1$, while $M$ and $M'$ 
are the mass eigenvalues of the baryon states
(here, we restrict our investigation to the elastic case with $M=M'$).
Conversely, $\sum_i \omega_i=\sum_i \sqrt{m_i^2+{\vec k}^2_i}$ 
is the invariant mass of the free incoming three-quark system, where $\vec k_i$
are internal momenta, and similarly for $\sum_i \omega_i'$ for the outgoing. 
Obviously, the normalization
factors so defined are all Lorentz invariant. Here, we further constrain the construction 
by adhering to additional global requirements for the EM amplitudes.

Charge normalization requires $x=3$ (for any $y$ values), 
because in~\fref{fig1}(a) only for this value
the proton charge is recovered at zero momentum transfer.
This result substantiates the findings in Ref.~\cite{Melde:2004qu}, where one
can see that only when the exponent takes the $x=3$ value, one can recover
a genuine one-body operator at zero momentum transfer.
If we 
impose that the EM-current amplitude is time-reversal invariant, then it has been
demonstrated (see Refs.~\cite{Durand:1962,Ernst:1960}) that the third component of 
${\vec F}_{\Sigma',\Sigma}$ has to vanish identically in the Breit frame. 
In~\fref{fig1}(b) we present the corresponding transition amplitude exhibiting an 
antisymmetric structure around the value
$y=\frac{1}{2}$. To obtain a vanishing amplitude the obvious choice clearly is 
the zero-crossing at $y=\frac{1}{2}$, leading to the symmetric (geometric) factor
\begin{equation}
{\cal N}_{\rm S}=
\left(\frac{M}{\sum_i{\omega_{i}}}\right)^{\frac{3}{2}}
\left(\frac{M'}{\sum_i{\omega'_{i}}}\right)^{\frac{3}{2}}\, .
\label{eq:geoN}
\end{equation}
However, the antisymmetric structure of the curves in~\fref{fig1}(b)
allows to consider also other, equally valid, PFSM normalizations. 
For example, physically valid constructions are
\begin{equation}
{\cal N}\left(z\right)=
\frac{1}{2}\left[
\left(\frac{M}{\sum_i{\omega_{i}}}\right)^{3z}
\left(\frac{M'}{\sum_i{\omega'_{i}}}\right)^{3\left(1-z\right)}
+
\left(\frac{M'}{\sum_i{\omega'_{i}}}\right)^{3z}
\left(\frac{M}{\sum_i{\omega_{i}}}\right)^{3\left(1-z\right)}
\right]
\, .
\label{eq:facgen}
\end{equation}
In Eq.~(\ref{eq:facgen}), one can sort out two special cases, namely ${\cal N}_{\rm S}$
for $z=0.5$ and ${\cal N}_{\rm ari}$ for $z=0$ (or $z=1$).
Here, ${\cal N}_{\rm S}$ is given by Eq.~(\ref{eq:geoN}), while ${\cal N}_{\rm ari}$ 
takes on the form of an  arithmetic combination
\begin{equation}
{\cal N}_{\rm ari}=
\frac{1}{2}\left[
\left(\frac{M}{\sum_i{\omega_{i}}}\right)^{3}+
\left(\frac{M'}{\sum_i{\omega'_{i}}}\right)^{3}
\right]\, .
\label{eq:arithfac}
\end{equation}
All forms implied by Eq.~(\ref{eq:facgen}) lead to the correct charge normalization 
and fulfill time-reversal invariance\footnote{ 
a linear combination of the type $a{\cal N}_{\rm S}+\left(1-a\right){\cal N}_{\rm ari}$
also is an allowed construction.}.
In the $Q^2\rightarrow 0$ limit all different expressions implied 
by Eq.~(\ref{eq:facgen}) lead to the same transition amplitudes.

In~\fref{fig2} we show the ratio of  magnetic nucleon to standard dipole 
form factors in comparison to experiment.
As one moves to higher $Q^2$ momenta, one observes a broadening of the
band between the curves ${\cal N}_{\rm S}$ and ${\cal N}_{\rm ari}$, 
indicating that the normalization factor effects mainly the higher momentum 
results. The predictions with ${\cal N}_{\rm S}$ 
(full line) exhibit a strong fall off above $Q^2=1$GeV$^2$ following approximately the lower
bounds of the experimental neutron data, but considerably underestimating the 
proton data. The dashed line, representing the results with ${\cal N}_{\rm ari}$,
exhibit an overpredicton of the experimental data. 
We also have performed a one-parameter "best fit" (denoted by the dash-dotted line)
to the experimental data of the dipole ratios $G_M^n/G_D$ and $G_M^p/G_D$ 
at $3$GeV$^2$ momentum transfer. 
The value $z=1/6$ in Eq.~(\ref{eq:facgen}) produces a theoretical curve that lies still below
the experimental data for the proton, while it is already above for the neutron. 
These reults are denoted as ${\cal N}_{\rm fit}$ and represented by the  
dash-dotted lines in Figs.~\ref{fig2},\ref{fig3}.
The two curves with ${\cal N}_{\rm geo}$ and ${\cal N}_{\rm ari}$
envelop the experimental data in most cases, while the curve with ${\cal N}_{\rm fit}$
lies in-between. 

In~\fref{fig3} we show the ratio of the electric and magnetic proton
form-factor (left) and the ratio of the electric proton form-factor to standard dipole 
parametrization (right). Again, the theoretical spread enlarges with increased 
momentum transfer.
The dash-dotted line follows the experimental values and is congruent to latest data with the 
asymmetric beam-target experiment by Jones et al.~\cite{Jones:2006kf}. Note that
this line was obtained with the same ${\cal N}_{\rm fit}$, which was determined previously.
To further decrease the theoretical spread, one needs to find additional conditions
that further constrain the construction of PFSM operators.
However, for the time being one could assume the band between the two curves 
as an estimate for the theoretical uncertainty of the PFSM construction of the 
nucleon EM form-factors.

\section{Summary and Conclusion
\label{label:Conclusion}
}
In this contribution we have reconsidered the construction of 
EM operators in the PFSM formalism. The spectator-like
nature of the construction implies that only one quark is explicitly coupled to 
the current operator,  but nevertheless the operator contains many-body effects.
Such many-body effects appear in the kinematics and in the occurrence of a 
normalization factor that is explicitly required by the theory. The specific expression
for the factor, however, is not completely determined since Poincar\'e
invariance and proton-charge normalization alone are not sufficient to
uniquely determine the current-operator. Time-reversal invariance produces
an important additional constraint that further restricts the possible choices,
but still allows for a residual indetermination. This spread is zero at
zero-momentum transfer and increases at higher $Q^2$-values.
We have estimated the corresponding theoretical uncertainty
and found that it remains reasonably small, at least for the EM form-factors of
the nucleon. We have finally determined an optimal choice
utilizing one single free parameter, obtaining results overall in good
agreement with the experimental data.
\section*{Acknowledgements}
The results reported in this contribution originated in a 
collaborative effort with L. Canton, W. Plessas and R.~F. Wagenbrunn.
The author is particularly thankful to L. Canton for his help in completing
this manuscript and to R.~F. Wagenbrunn for his expertise in the numerics.
The author is grateful to the INFN Sezione di Padova and 
the University of Padova for their hospitality. 
This work was supported by the Austrian Science Fund 
(Projects P16945 and  P19035). Finally, the author would like to thank
the organizers for accepting this contribution to the Cortona 
meeting.
\bibliographystyle{ws-procs9x6}
%\bibliography{Current}

\end{document}